\documentclass{article}
\usepackage{sw20lart}

\input{tcilatex}
\begin{document}

\title{An asymptotic form of the reciprocity theorem with applications in x-ray
scattering}
\author{Ariel Caticha \\
Department of Physics, University at Albany-SUNY, \\
Albany, NY 12222, USA. \\
ariel@cnsvax.albany.edu}
\date{}
\maketitle

\begin{abstract}
The emission of electromagnetic waves from a source within or near a
non-trivial medium (with or without boundaries, crystalline or amorphous,
with inhomogeneities, absorption and so on) is sometimes studied using the
reciprocity principle. This is a variation of the method of Green's
functions. If one is only interested in the asymptotic radiation fields the
generality of these methods may actually be a shortcoming: obtaining
expressions valid for the uninteresting near fields is not just a wasted
effort but may be prohibitively difficult. In this work we obtain a modified
form the reciprocity principle which gives the asymptotic radiation field
directly. The method may be used to obtain the radiation from a prescribed
source, and also to study scattering problems. To illustrate the power of
the method we study a few pedagogical examples and then, as a more
challenging application we tackle two related problems. We calculate the
specular reflection of x rays by a rough surface and by a smoothly graded
surface taking polarization effects into account. In conventional treatments
of reflection x rays are treated as scalar waves, polarization effects are
neglected. This is a good approximation at grazing incidence but becomes
increasingly questionable for soft x rays and UV at higher incidence angles.

PACs: 61.10.Dp, 61.10.Kw, 03.50.De
\end{abstract}

\section{Introduction}

The principle of reciprocity can be traced to Helmholtz in the field of
acoustics. It states that everything else being equal the amplitude of a
wave at a point $A$ due to a source at point $B$ is equal to the amplitude
at $B$ due to a source at $A$. With its extension to electromagnetic waves
by Lorentz \cite{Lorentz05} and later to quantum mechanical amplitudes \cite
{Blatt79}, the applicability to all sorts of fields was made manifest.
Nowadays the principle is regarded as a symmetry of Green's functions when
the source point and the field point are reversed. This symmetry is actually
quite general. As shown in \cite{Bilhorn64} the conditions of time-reversal
invariance and hermiticity of the Hamiltonian are sufficient to guarantee
reciprocity, but they are not necessary; in fact, reciprocity holds even in
the presence of complex absorbing potentials. In the case of electromagnetic
waves the only requirement is that the material medium be linear and
described by symmetric permittivity and permeability tensors \cite{Landau84}%
\cite{Kong90}. This excludes plasmas and ferrite media in the presence of
magnetic fields.

In the field of x-ray optics the principle was used by von Laue \cite{Laue35}
to explain the diffraction patterns generated by sources within the crystal,
the so-called Kossel lines \cite{Kossel35}. More recently there has been a
widespread recognition that these interference patterns contain information
not just about intensities but also about phases and can be thought of as
holographic records from which real space images of the location of the
internal sources can be reconstructed. Thus, under the modern name of `x-ray
holography' there has been a considerable revival of interest in this
subject \cite{Tegze91}.

However, powerful as it is, the usual formulation of the reciprocity
principle suffers from a rather serious drawback: it refers to the exchange
of source and field \emph{points}. As a consequence, a careful application
of the principle requires one to consider the emission of spherical waves
which in crystalline media or even in the mere presence of plane boundaries,
can be surprisingly difficult (recall \emph{e.g.} studying the radiation by
an antenna in the vicinity of the conducting surface of the Earth \cite
{Stratton41} or of layered media \cite{Chew90}). Furthermore, one is
typically interested in the asymptotic radiation fields so the relevant
exchange should involve a source point \emph{here} with a field point at 
\emph{infinity}.

These technical difficulties have not deterred the users of reciprocity from
using the principle to make valuable predictions, but a high price has been
paid. The required asymptotic limits are usually taken \emph{verbally} and
no accounts are given of where and how spherical waves are replaced by plane
waves. Such sleights of hand, because skillfully performed, have not lead to
wrong results, but intensities are predicted only up to undetermined
proportionality factors and this excludes applications to classes of
problems where absolute intensities are needed. Moreover one is left with
the uneasy feeling that the validity of the predictions is justified mostly
on the purely pragmatic grounds that for the problem at hand they seem to
work which, again, limits applications to problems that are already familiar.

The main goal of this paper (section 2) is to obtain a modified form of the
reciprocity theorem that gives the asymptotic radiation fields directly and
that accommodates plane waves and both point and extended sources in a
natural way. Remarkably the resulting expressions, which include all the
relevant proportionality factors and yield absolute, not just relative
intensities, are very simple.

For many problems the Asymptotic Reciprocity Theorem (ART) obtained here
represents an improvement not only over the usual form of the reciprocity
theorem but also over the method of Green's functions. Computing the Green's
function requires solving a boundary value problems for spherical waves in
the presence of plane boundaries and/or periodic media; this may well be an
intractably difficult problem. Furthermore, a considerable effort is wasted
by first obtaining both near fields and far fields and then discarding the
uninteresting near fields. The ART is a shortcut that discards the near
fields before, rather than after they are computed.

To illustrate the power of the method we consider several applications. The
first three (section 3) are brief pedagogical examples of increasing
complexity. First the ART is used to calculate the fields radiated by an
arbitrary prescribed source in vacuum; next as an application to scattering
problems we reproduce the kinematical theory of diffraction by crystals. The
third example, the radiation by a current located near a plane dielectric
boundary, is straightforward when the ART is used but not if other methods
are used. One must emphasize that what is new in these examples are not the
results, but the method; the first two are standard textbook material, a
special case of the third is treated in \cite{Stratton41}. As a more
involved application of the ART, in section 4 we combine ideas from the
three previous examples to study two other related scattering problems, the
specular reflection of polarized x rays by a rough surface and by a
continuously graded surface.

The technique of the grazing-incidence reflection of x-rays has received
considerable attention \cite{Beckmann87}-\cite{Caticha95} from both the
theoretical and the experimental sides as a means to obtain structural
information about surfaces. The effect of surface roughness on the
reflection is taken into account by multiplying the Fresnel reflectivity of
an ideal sharp and planar surface by a ``static Debye-Waller'' factor. The
problem is to calculate this corrective factor. The calculation has been
carried out in several different approximations. The Rayleigh or Born
approximation \cite{Beckmann87} is satisfactory for rough surfaces with long
lateral correlation lengths but for x-rays the situations of interest
generally involve short lateral correlation lengths. Here other
approximations such as the distorted-wave Born approximation \cite{Sinha88}%
\cite{Dietrich95} and the Nevot-Croce approximation \cite{NevotCroce80} are
used. For variations and interpolations between these two methods see \cite
{Pynn92}, and for a generalization to surfaces with non-Gaussian roughness
and to graded interfaces of arbitrary profile see \cite{Caticha95}. In these
treatments (\cite{Dietrich95} is an exception) the x rays are treated as
scalar waves. One expects this approximation to hold at grazing incidence
but at higher incidence angles (\emph{e.g.}, for soft x rays) its validity
becomes increasingly questionable. Using a modified first Born approximation
Dietrich and Haase \cite{Dietrich95} took the vector character of the x rays
into account but they point out that the validity of their approximation is
not in general easy to assess and they restrict themselves to studying
special interface profiles.

In section 4 we study this problem using a different approximation; we use
the ART to develop approximations of the Nevot-Croce type \cite{Caticha95}.
There is, of course, a trivial polarization dependence that is already
described by Fresnel formulas for the reflectivity of the ideal flat step
surface. The question we address here is whether the ``static Debye-Waller''
factor shows any additional dependence on polarization. The final result is
remarkably simple: the ``static Debye-Waller'' factor for the specularly
reflected vector waves is the same for both polarizations and coincides with
that for scalar waves. Finally, some brief concluding remarks are collected
in section 5.

\section{The reciprocity theorem and its asymptotic form}

We wish to calculate the asymptotic radiation fields $\vec{E}$ and $\vec{H}$
generated by a prescribed current $\vec{J}\left( t,\vec{r}\right) $ located
near or within a linear medium,

\[
D_i=\varepsilon _{ij}E_j\,\,\,\,\,\,\,\,\text{and}\,\,\,\,\,\,\,B_i=\mu
_{ij}H_j\text{ .} 
\]
We will assume that the tensors $\varepsilon _{ij}\left( \vec{r}\right) $
and $\mu _{ij}\left( \vec{r}\right) $ are symmetric, but otherwise the
situation remains quite general, the medium may have an irregular shape, or
be inhomogeneous, crystalline or amorphous, absorbing, dispersive, etc.

As in the usual deduction of the reciprocity theorem (see \emph{e.g.}, \cite
{Landau84}), we consider a second set of fields $\vec{E}_c$ and $\vec{H}_c$,
which we will call the ``connecting fields'', generated by a source $\vec{J}%
_c$ (see fig.\ref{fig1}a). For simplicity we will also assume that all
fields and sources are monochromatic $\vec{E}=\vec{E}(\vec{r})e^{-i\omega t}$%
, $\vec{J}=\vec{J}(\vec{r})e^{-i\omega t}$, etc. For linear media this is
not a restriction.

\FRAME{ftFU}{4.4218in}{2.2727in}{0pt}{\Qcb{(a) In the usual form of the
reciprocity theorem the surface $S$ encloses the medium, and all sources.
(b) For the asymptotic form of the reciprocity theorem the connecting field $%
\vec{E}_c$ is a radiation field, its source lies outside the surface $%
S=S_{+}+S^{^{\prime }}.$ }}{\Qlb{fig1}}{Figure }{\special{language
"Scientific Word";type "GRAPHIC";maintain-aspect-ratio TRUE;display
"USEDEF";valid_file "T";width 4.4218in;height 2.2727in;depth
0pt;original-width 284.8125pt;original-height 220pt;cropleft "0";croptop
"0.7794";cropright "1";cropbottom "0.1175";tempfilename
'C:/A-PAPERS/RecipThm/submit/figure1.eps';tempfile-properties "XNP";}}

From Maxwell's equations 
\begin{equation}
\nabla \times \vec{E}=iK\vec{B}\text{\qquad and\qquad }\nabla \times \vec{H}%
=-iK\vec{D}+\frac{4\pi }c\vec{J},  \label{Maxwell}
\end{equation}
where $K\equiv \omega /c$, one easily obtains the following identity 
\[
\nabla \cdot \left( \vec{E}\times \vec{H}_c-\vec{E}_c\times \vec{H}\right) =%
\frac{4\pi }c\left( \vec{E}_c\cdot \vec{J}-\vec{E}\cdot \vec{J}_c\right) , 
\]
which, on integrating over a large volume $V$ bounded by the surface $S$,
can be rewritten as 
\begin{equation}
\int_S\left( \vec{E}\times \vec{H}_c-\vec{E}_c\times \vec{H}\right) \cdot d%
\vec{s}=\frac{4\pi }c\int_V\left( \vec{E}_c\cdot \vec{J}-\vec{E}\cdot \vec{J}%
_c\right) \,dv.  \label{mainidentity}
\end{equation}
This expression simplifies if one deals with point sources. For example,
consider oscillating point dipoles $\vec{p}_oe^{-i\omega t}$ and $\vec{p}%
_ce^{-i\omega t}$, located at $\vec{r}_o$ and $\vec{r}_c$ respectively. The
current density $\vec{J}$ is given by $\vec{J}=-i\omega \vec{p}_o\delta
\left( \vec{r}-\vec{r}_o\right) e^{-i\omega t}$ and $\vec{J}_c$ is given by
an analogous expression. Further simplification is achieved if one assumes
that the surface $S$ is so remote that the surface integral is negligibly
small \cite{footnote1}, then 
\begin{equation}
\vec{E}_c\left( \vec{r}_o\right) \cdot \vec{p}_o=\vec{E}\left( \vec{r}%
_c\right) \cdot \vec{p}_c\text{ .}  \label{usualRT}
\end{equation}
This is the usual form of the reciprocity theorem; it says that if we know $%
\vec{E}_c$ at the location of $\vec{p}_o$ we can calculate $\vec{E}$ at the
location of $\vec{p}_c$. This elegant result takes us a long way toward a
final answer for $\vec{E}$, but the remaining problem of calculating $\vec{E}%
_c$, that is, the calculation of how the spherical wave generated by $\vec{p}%
_c$ is scattered by the medium, can still be too difficult.

A more useful version of the theorem can be obtained once one realizes that
the connecting field is merely a tool that codifies information about the
influence of the non-trivial medium. Above, the field $\vec{E}_c$ has been
introduced by first specifying a source $\vec{J}_c$, but clearly this is an
unnecessary additional complication. In fact, since the most convenient $%
\vec{J}_c$ is that which results in the simplest $\vec{E}_c$ it is best to
focus attention directly on the field rather than its source. Thus we move $%
\vec{J}_c$ outside the surface $S$, to infinity (see fig.\ref{fig1}b) so
that throughout the volume $V$ the connecting field $\vec{E}_c$ is a pure
radiation field. Furthermore, let the surface $S$ itself be so distant that
on $S$ itself both $\vec{E}$ and $\vec{E}_c$ are \emph{vacuum} radiation
fields. Then 
\begin{equation}
\int_S\left( \vec{E}\times (\nabla \times \vec{E}_c)-\vec{E}_c\times (\nabla
\times \vec{E})\right) \cdot d\vec{s}=\frac{4\pi iK}c\int_V\vec{E}_c\cdot 
\vec{J}\,dv.  \label{art1}
\end{equation}
At this point it is not yet clear that this form of the reciprocity theorem
is simpler than eq. (\ref{usualRT}) but one remarkable feature can already
be seen: eq. (\ref{art1}) relates the field $\vec{E}$ at a distant surface $%
S $ to its source $\vec{J}$ within a nontrivial medium \emph{without} having
to calculate $\vec{E}$ in the vicinity of $\vec{J}$. The ``connection''
between the distant radiation field $\vec{E}$ and its source $\vec{J}$ is
achieved through the much simpler (i.e., hopefully calculable)
``connecting'' field $\vec{E}_c$.

To bring eq. (\ref{art1}) into a form that is manifestly simpler than (\ref
{usualRT}) the surface $S$ is chosen as a cube with edges of length $%
L\rightarrow \infty $. In fig.\ref{fig1}b the upper face, defined by a
constant $z$ coordinate, $z=z_{+}$, has been singled out as $S_{+}$, the
remaining seven faces are denoted $S^{^{\prime }}$.

On the upper face $S_{+}$ we write the field $\vec{E}$ as a superposition of
outgoing plane waves of wave vector $\vec{k}$ satisfying $\vec{k}\cdot \vec{k%
}=\omega ^2/c^2=K^2$ and $k_z>0$,

\begin{equation}
\vec{E}\left( \vec{r}\right) =\int\limits_{k_z>0}\frac{d^3k}{\left( 2\pi
\right) ^3}\,2\pi \delta \left( k-K\right) \,\vec{E}(\vec{k})\,e^{i\vec{k}%
\cdot \vec{r}}\,,
\end{equation}
where, in a self-explanatory notation, $\vec{k}\cdot \vec{r}=\vec{k}_{\perp
}\cdot \vec{r}_{\perp }+k_zz_{+}$. It is here, by the very act of writing $%
\vec{E}$ in this form, that the asymptotic limit of discarding near fields
is being taken. For $z<<z_{+}$ additional terms describing the near fields
should be included.

The integral over $dk_z$ is most easily done using 
\begin{equation}
\delta \left( k-K\right) =\frac K{k_z}\left[ \delta \left( k_z-\sqrt{%
K^2-k_{\perp }^2}\right) -\delta \left( k_z+\sqrt{K^2-k_{\perp }^2}\right)
\right] .
\end{equation}
The result is 
\begin{equation}
\vec{E}\left( \vec{r}\right) =\int \frac{d^2k_{\perp }}{\left( 2\pi \right)
^2}\,\frac K{k_z}\,\vec{E}(\vec{k})\,e^{i\vec{k}\cdot \vec{r}},
\label{RadField}
\end{equation}
where $k_z=+\sqrt{K^2-k_{\perp }^2}$.

The choice of the connecting field $\vec{E}_c$ is dictated purely by
convenience. A particularly good choice for $\vec{E}_c$ is the superposition
of an incoming plane wave of unit amplitude (we use $\symbol{94}$ to denote
vectors of unit length) and wave vector $\vec{k}_c$, with $\vec{k}_c\cdot 
\vec{k}_c=K^2$ and $k_{cz}<0$, plus all the waves scattered by the medium, 
\begin{equation}
\vec{E}_c\left( \vec{r}\right) =\hat{e}_c\,e^{i\vec{k}_c\cdot \vec{r}}+\vec{E%
}_c^{^{\prime }}\left( \vec{r}\right) .  \label{ConnField1}
\end{equation}
On $S_{+}$, the scattered field $\vec{E}_c^{^{\prime }}$ is a superposition
of outgoing plane waves and is given also in a form analogous to eq. (\ref
{RadField}), 
\begin{equation}
\vec{E}_c^{^{\prime }}\left( \vec{r}\right) =\int \frac{d^2k_{\perp
}^{^{\prime }}}{\left( 2\pi \right) ^2}\,\frac K{k_z^{^{\prime }}}\,\vec{E}%
_c^{^{\prime }}(\vec{k}^{^{\prime }})\,e^{i\vec{k}^{^{\prime }}\cdot \vec{r}%
},  \label{ConnField2}
\end{equation}
with $\vec{k}^{^{\prime }}\cdot \vec{r}=\vec{k}_{\perp }^{^{\prime }}\cdot 
\vec{r}_{\perp }+k_z^{^{\prime }}z_{+}$ and $k_z^{^{\prime }}=+\sqrt{%
K^2-k_{\perp }^{^{\prime }2}}$.

Now we are ready to calculate the surface integral on the left hand side of
eq. (\ref{art1}). Substituting (\ref{ConnField1}) into (\ref{art1}) the
integral over $S_{+}$ separates into two terms, one due to the incoming
plane wave $\hat{e}_c\,e^{i\vec{k}_c\cdot \vec{r}}$, and the other due to
the scattered waves $\vec{E}_c^{^{\prime }}\left( \vec{r}\right) $. The
first term is 
\begin{equation}
I_1=\int_{S_{+}}dx\,dy\,\hat{e}_z\cdot \left( \vec{E}\times (i\vec{k}%
_c\times \hat{e}_c)-\hat{e}_c\times (\nabla \times \vec{E})\right) e^{i\vec{k%
}_c\cdot \vec{r}}\text{,}
\end{equation}
and substituting (\ref{RadField}) its evaluation is straightforward. The
integral over $dx\,dy$ yields $(2\pi )^2\delta (\vec{k}_{\perp }+\vec{k}%
_{c\perp })$. Since $k^2=k_c^2=K^2$, $k_{cz}<0$ and $k_z>0$ this implies
that the plane waves superposed in (\ref{RadField}) yield a vanishing
contribution except when $\vec{k}=-\vec{k}_c$. Thus, 
\begin{equation}
I_1=-2iK\,\hat{e}_c\cdot \vec{E}(-\vec{k}_c)  \label{I1}
\end{equation}

The contribution of the scattered waves $\vec{E}_c^{^{\prime }}\left( \vec{r}%
\right) $, 
\begin{equation}
I_2=\int_{S_{+}}dx\,dy\,\hat{e}_z\cdot \left( \vec{E}\times (\nabla \times 
\vec{E}_c^{^{\prime }})-\vec{E}_c^{^{\prime }}\times (\nabla \times \vec{E}%
)\right) ,
\end{equation}
is calculated in a similar way. Substitute (\ref{RadField}) and (\ref
{ConnField2}) and integrate over $dx\,dy$ to obtain a delta function. This
eliminates all Fourier components except those with $\vec{k}_{\perp }=-\vec{k%
}_{\perp }^{^{\prime }}$. Since $k^2=k^{^{\prime }2}=K^2$, and both $%
k_z,k_z^{^{\prime }}>0$ this implies $k_z=$ $k_z^{^{\prime }}>0$. Thus, 
\begin{equation}
I_2=\int \frac{d^2k_{\perp }^{^{\prime }}}{\left( 2\pi \right) ^2}\,\frac{K^2%
}{k_z^{^{\prime 2}}}\,\hat{e}_z\cdot \left[ \vec{E}(\vec{k})\times \left( i%
\vec{k}^{^{\prime }}\times \vec{E}_c^{^{\prime }}(\vec{k}^{^{\prime
}})\right) -\vec{E}_c^{^{\prime }}(\vec{k}^{^{\prime }})\times \left( i\vec{k%
}\times \vec{E}(\vec{k})\right) \right] ,
\end{equation}
where $\vec{k}=-\vec{k}^{^{\prime }}+2k_z^{^{\prime }}\hat{e}_z$. Further
manipulation using $\vec{k}^{^{\prime }}\cdot \vec{E}_c^{^{\prime }}(\vec{k}%
^{^{\prime }})=\vec{k}\cdot \vec{E}(\vec{k})=0$ gives $\hat{e}_z\cdot
[\cdots ]=0$, so that 
\begin{equation}
I_2=0.  \label{I2}
\end{equation}

According to eq. (\ref{I1}) and (\ref{I2}) the only contributions to the
surface integral over the distant plane $S_{+}$ come from products of
outgoing with incoming waves. Products of two outgoing waves yield vanishing
contributions. This result applies also to the remaining seven faces of the
cube $S$. Since on each of these faces there are only outgoing waves we find
that the integral over $S^{^{\prime }}$ makes no contribution to the left
hand side of (\ref{art1}). Incidentally, this argument completes our
previously unfinished deduction of the usual form of the reciprocity
theorem, eq. (\ref{usualRT}): if both sources $\vec{J}$ and $\vec{J}_c$ are
internal to the surface $S$ the surface integral in eq. (\ref{mainidentity})
vanishes because it only involves products of outgoing waves.

Substituting (\ref{I1}) into (\ref{art1}) leads us to the main result of
this paper, the asymptotic reciprocity theorem, 
\begin{equation}
\widehat{e}\cdot \vec{E}(\vec{k})=-\frac{2\pi }c\int_V\vec{E}_c\cdot \vec{J}%
\,dv.  \label{art}
\end{equation}
In words:

\emph{The field }$\vec{E}(\vec{k})$\emph{\ radiated in a direction }$\vec{k}$%
\emph{\ with a certain polarization }$\widehat{e}$\emph{\ is }$-2\pi /c$%
\emph{\ times the ``component'' of the source }$\vec{J}(\vec{r})$\emph{\
``along'' a connecting field }$\vec{E}_c(\vec{r})$\emph{\ with incoming wave
vector }$\vec{k}_c=-\vec{k}$\emph{\ and polarization }$\widehat{e}_c=%
\widehat{e}$.

Typically one is interested in the intensity radiated into a solid angle $%
d\Omega $; since the amplitude $\vec{E}(\vec{k})$ that appears in (\ref
{RadField}) and (\ref{art}) is not quite the Fourier transform of $\vec{E}(%
\vec{r})$ it may be useful to derive an explicit expression for $dW/d\Omega $%
. The total power radiated through the plane $S_{+}$ is given by the flux of
the time-averaged Poynting vector, $\frac c{8\pi }\func{Re}\,[\vec{E}\times 
\vec{B}^{*}]$, 
\begin{equation}
W=\int d^2x_{\perp }\frac c{8\pi }\func{Re}\,[\vec{E}\times \vec{B}%
^{*}]\cdot \hat{e}_z=\int d\Omega \,\frac{dW}{d\Omega }
\end{equation}
Using (\ref{RadField}) and $d^2k_{\perp }=k_{\perp }dk_{\perp }d\phi
=Kk_zd\Omega $ (where $\phi $ is the usual azimuthal angle about the $z$
axis) we get 
\begin{equation}
W=\frac c{8\pi }\int \frac{d^2k_{\perp }}{\left( 2\pi \right) ^2}\frac K{k_z}%
\vec{E}(\vec{k})\cdot \vec{E}^{*}(\vec{k}),
\end{equation}
so that 
\begin{equation}
\frac{dW}{d\Omega }=\frac c{8\pi }\left( \frac K{2\pi }\right) ^2\,\vec{E}(%
\vec{k})\cdot \vec{E}^{*}(\vec{k}).  \label{Power}
\end{equation}
\ In the next section we offer a few illustrative examples of the ART in
action.

\section{Some simple examples}

The ART, eq. (\ref{art}), holds for an arbitrary linear medium. In
particular it holds if the medium is vacuum. Our first trivial example is
the radiation by a prescribed current in vacuum. Next, to show that the ART
can be used to study scattering problems we deal with another equally
trivial example, the kinematical theory of diffraction by crystals. The
third example, the radiation by currents located near a dielectric boundary,
is also straightforward. What is remarkable here is the ease with which the
results are obtained compared to conventional methods \cite{Stratton41}\cite
{Chew90}.

\subsection{Radiation in vacuum}

In this case the connecting field is just an incoming plane wave, $\vec{E}%
_c\left( \vec{r}\right) =\hat{e}_c\,e^{i\vec{k}_c\cdot \vec{r}}$. The ART,
eq. (\ref{art}), gives the radiated field with polarization $\hat{e}=\hat{e}%
_c$ as 
\begin{equation}
\widehat{e}\cdot \vec{E}(\vec{k})=-\frac{2\pi }c\int_V\vec{J}(\vec{r})\cdot 
\hat{e}\,e^{-i\vec{k}\cdot \vec{r}}\,dv=-\frac{2\pi }c\,\hat{e}\cdot \vec{J}(%
\vec{k}),
\end{equation}
so that 
\begin{equation}
\vec{E}(\vec{k})=\frac{2\pi }c\hat{k}\times (\hat{k}\times \vec{J}(\vec{k})).
\end{equation}
The radiated power, eq. (\ref{Power}), is 
\begin{equation}
\frac{dW}{d\Omega }=\frac{K^2}{8\pi c}\left| \hat{k}\times (\hat{k}\times 
\vec{J}(\vec{k}))\right| ^2,
\end{equation}
as expected. (For radiation by a point dipole just substitute $\vec{J}(\vec{k%
})=-icK\vec{p}$.)

\subsection{Bragg diffraction}

Consider a crystal described by its dielectric susceptibility $\chi (\vec{r}%
) $ \cite{footnote2} which for x rays is quite small (typically about $%
10^{-5}$ or less). An incident plane wave $\vec{E}_oe^{i\vec{k}_o\cdot \vec{r%
}}$ induces a current 
\begin{equation}
\vec{J}(\vec{r})=-i\omega \vec{P}(\vec{r})=\frac{-i\omega }{4\pi }\chi (\vec{%
r})\vec{E}_oe^{i\vec{k}_o\cdot \vec{r}},
\end{equation}
which radiates. The connecting field needed to calculate this radiation is a
simple incoming plane wave, $\vec{E}_c\left( \vec{r}\right) =\hat{e}_c\,e^{i%
\vec{k}_c\cdot \vec{r}}$, and the ART, eq. (\ref{art}), gives the radiated
field as 
\begin{equation}
\vec{E}(\vec{k})=-\frac{i\omega }{2c}\hat{k}\times (\hat{k}\times \vec{E}%
_o)\,\chi (\vec{k}-\vec{k}_o).
\end{equation}
The scattered field is proportional to the Fourier transform of the
susceptibility of the medium; for a periodic medium this is Bragg
diffraction.

\subsection{Radiation in the vicinity of a reflecting surface}

Consider a current $\vec{J}_{in}(\vec{r})$ located within a uniform medium
with dielectric susceptibility $\chi _0$ occupying the region $z<0$ (see
fig. \ref{Fig2}). To calculate the radiation in the direction $\vec{k}$ with
polarization $\hat{e}$ we choose as connecting field an incoming plane wave
with wave vector $\vec{k}_c=-\vec{k}$ and unit amplitude $\hat{e}_c=\hat{e}$
plus the corresponding reflected and transmitted waves, 
\begin{equation}
\vec{E}_c\left( \vec{r}\right) =\left\{ 
\begin{array}{ll}
\hat{e}_c\,e^{i\vec{k}_c\cdot \vec{r}}+\vec{\varepsilon}_{cr}\,e^{i\vec{k}%
_{cr}\cdot \vec{r}}, & \text{for\quad }z>0 \\ 
\vec{\varepsilon}_{ct}\,e^{i\vec{k}_{ct}\cdot \vec{r}}, & \text{for\quad }z<0
\end{array}
\right.  \label{ConnE}
\end{equation}
The various wave vectors are given by 
\begin{equation}
\vec{k}_c=-K\cos \theta \,\hat{e}_x-q\hat{e}_z=-\vec{k}\,,
\end{equation}
\begin{equation}
\vec{k}_{cr}=-K\cos \theta \,\hat{e}_x+q\hat{e}_z\,,
\end{equation}
\begin{equation}
\vec{k}_{ct}=-K\cos \theta \,\hat{e}_x-\bar{q}\hat{e}_z\,,
\end{equation}
where $K=\omega /c$, and the normal components $q$ and $\bar{q}$ are given
by 
\begin{equation}
q=K\sin \theta \quad \text{and\quad }\bar{q}=K\left( \sin ^2\theta +\chi
_0\right) ^{1/2}\,.
\end{equation}
The amplitudes $\vec{\varepsilon}_{cr}$ and $\vec{\varepsilon}_{ct}$ of the
reflected and transmitted waves are given by the Fresnel expressions 
\begin{equation}
\vec{\varepsilon}_{cr}=r_s\hat{e}_{cr}\quad \text{where\quad }r_s=\frac{q-%
\bar{q}}{q+\bar{q}}\,\frac{\hat{e}_{cr}\cdot \hat{e}_{ct}}{\hat{e}_c\cdot 
\hat{e}_{ct}},
\end{equation}
and 
\begin{equation}
\vec{\varepsilon}_{ct}=t_s\hat{e}_{ct}\quad \text{where\quad }t_s=\frac{2q}{%
q+\bar{q}}\,\frac 1{\hat{e}_c\cdot \hat{e}_{ct}}.  \label{ect}
\end{equation}
($\hat{e}_{cr}$ and $\hat{e}_{ct}$ are unit vectors describing the
polarization of the specular reflected and transmitted waves.)

\FRAME{ftbpFU}{3.7403in}{2.9412in}{0pt}{\Qcb{The connecting field for
radiation in the presence of a reflecting medium includes reflected and
transmitted waves. Here the source $\vec{J}_{in}$ is shown within the medium
($z<0$).}}{\Qlb{Fig2}}{Figure }{\special{language "Scientific Word";type
"GRAPHIC";maintain-aspect-ratio TRUE;display "USEDEF";valid_file "T";width
3.7403in;height 2.9412in;depth 0pt;original-width 284.8125pt;original-height
220pt;cropleft "0.1748";croptop "0.7736";cropright "0.8251";cropbottom
"0.1131";tempfilename 'C:/A-PAPERS/RecipThm/submit/figure2.eps';tempfile-properties
"XP";}}

Then for a source $\vec{J}_{in}(\vec{r})$ located within the medium, the
ART, eq.(\ref{art}), gives the radiated field as 
\begin{equation}
\widehat{e}\cdot \vec{E}(\vec{k})=-\frac{2\pi }c\int_{z<0}\vec{J}_{in}(\vec{r%
})\cdot \vec{\varepsilon}_{ct}\,e^{i\vec{k}_{ct}\cdot \vec{r}}\,dv\,.
\label{Erad1}
\end{equation}
On the other hand, had the source $\vec{J}_{out}(\vec{r})$ been located
outside the dielectric medium ($z>0$) the corresponding radiated field would
be 
\begin{equation}
\widehat{e}\cdot \vec{E}(\vec{k})=-\frac{2\pi }c\int_{z>0}\vec{J}_{out}(\vec{%
r})\cdot \left( \hat{e}_c\,e^{i\vec{k}_c\cdot \vec{r}}+\vec{\varepsilon}%
_{cr}\,e^{i\vec{k}_{cr}\cdot \vec{r}}\right) \,dv\,.  \label{Erad2}
\end{equation}
For an oscillating dipole on the $z$ axis, $\vec{J}(\vec{r})=-i\omega \vec{p}%
\delta \left( \vec{r}-z_p\hat{e}_z\right) $, eq.(\ref{Erad1}) and (\ref
{Erad2}) give 
\begin{equation}
\widehat{e}\cdot \vec{E}(\vec{k})=\left\{ 
\begin{array}{ll}
2\pi iK\left( \vec{p}\cdot \hat{e}\,e^{-iqz_p}+\vec{p}\cdot \hat{e}%
_{cr}r_s\,e^{iqz_p}\right) & \text{if\quad }z_p>0 \\ 
2\pi iK\,\vec{p}\cdot \hat{e}_{ct}t_s\,e^{-i\bar{q}z_p} & \text{if\quad }%
z_p<0.
\end{array}
\right.
\end{equation}
The power radiated with polarization $\hat{e}$, eq.(\ref{Power}), is 
\begin{equation}
\frac{dW}{d\Omega }=\left[ \frac c{8\pi }K^4\left( \hat{e}\cdot \vec{p}%
\right) ^2\right] \left| 1+\frac{\hat{e}_{cr}\cdot \vec{p}}{\hat{e}\cdot 
\vec{p}}r_s\,e^{2iqz_p}\right| ^2\quad \text{for\quad }z_p>0\,,
\end{equation}
and 
\begin{equation}
\frac{dW}{d\Omega }=\left[ \frac c{8\pi }K^4\left( \hat{e}\cdot \vec{p}%
\right) ^2\right] \left| \frac{\hat{e}_{ct}\cdot \vec{p}}{\hat{e}\cdot \vec{p%
}}t_s\,e^{-i\bar{q}z_p}\right| ^2\quad \text{for\quad }z_p<0\,.
\end{equation}
In these two expressions we can recognize the first factor (in square
brackets) as the power radiated by a dipole in vacuum. The second factor
accounts for the presence of the dielectric medium.

\section{Specular reflection of polarized x rays}

In this section ideas from the three previous examples are combined to study
two similar and considerably more involved scattering problems, the specular
reflection of polarized x rays by a rough surface and by graded interfaces.
We show that within approximations of the Nevot-Croce type grading and
roughness affect the specular reflectivity in a manner that is independent
of the polarization of the incident radiation.

\subsection{Reflection by rough surfaces}

The dielectric susceptibility $\chi (\vec{r})$ that describes the rough
surface from which we wish to scatter x rays is given by 
\begin{equation}
\chi \left( x,y,z\right) =\left\{ 
\begin{array}{ll}
0 & \text{for\quad }z>\zeta (x,y) \\ 
\chi _0 & \text{for\quad }z<\zeta (x,y)
\end{array}
\right.
\end{equation}
where the height $\zeta (x,y)$, is a Gaussian random variable with zero
mean, $\left\langle \zeta \right\rangle =0$, and variance $\left\langle
\zeta ^2\right\rangle =\sigma ^2$ (see fig. \ref{fig3}).

To apply the ART it is convenient to rewrite $\chi \left( \vec{r}\right) $
as 
\begin{equation}
\chi \left( \vec{r}\right) =\chi _s\left( \vec{r}\right) +\delta \chi \left( 
\vec{r}\right) ,
\end{equation}
where $\chi _s\left( \vec{r}\right) $ represents a medium with an ideally
flat surface at $z_0$, 
\begin{equation}
\chi _s\left( \vec{r}\right) =\left\{ 
\begin{array}{ll}
0 & \text{for\quad }z>z_0 \\ 
\chi _0 & \text{for\quad }z<z_0
\end{array}
\right.
\end{equation}
and $\delta \chi \left( \vec{r}\right) $ represents the roughness.

Let $\hat{e}_0\,e^{i\vec{k}_0\cdot \vec{r}}$ be the incident field. The
total scattered field $\vec{\varepsilon}(\vec{r})$ includes the wave $\vec{%
\varepsilon}_s(\vec{r})$ specularly reflected by the step $\chi _s\left( 
\vec{r}\right) $ plus waves $\delta \vec{\varepsilon}\left( \vec{r}\right) $
scattered by $\delta \chi \left( \vec{r}\right) $ 
\begin{equation}
\vec{\varepsilon}(\vec{r})=\vec{\varepsilon}_s(\vec{r})+\delta \vec{%
\varepsilon}\left( \vec{r}\right) .  \label{scattE}
\end{equation}
The first term on the right is 
\begin{equation}
\vec{\varepsilon}_s(\vec{r})=\hat{e}_r\,r_se^{-2iqz_0}e^{i\vec{k}_r\cdot 
\vec{r}},  \label{epss}
\end{equation}
where 
\begin{equation}
r_s=\frac{q-\bar{q}}{q+\bar{q}}\,\frac{\hat{e}_r\cdot \hat{e}_t}{\hat{e}%
_0\cdot \hat{e}_t}
\end{equation}
($\hat{e}_r$ and $\hat{e}_t$ are unit vectors describing the polarization of
the specular reflected and transmitted waves). The second contribution in
eq.(\ref{scattE}), the field $\delta \vec{\varepsilon}\left( \vec{r}\right) $
includes a specular component plus diffusely scattered and evanescent waves, 
\begin{equation}
\delta \vec{\varepsilon}\left( \vec{r}\right) =\delta \varepsilon _r\,\hat{e}%
_re^{i\vec{k}_r\cdot \vec{r}}+\delta \vec{\varepsilon}_d(\vec{r}).
\label{delta eps}
\end{equation}
Using eq.(\ref{RadField}) this may be written as 
\begin{equation}
\delta \vec{\varepsilon}\left( \vec{r}\right) =\int \frac{d^2k_{\perp
}^{^{\prime }}}{\left( 2\pi \right) ^2}\,\frac K{k_z^{^{\prime }}}\delta 
\vec{\varepsilon}(\vec{k}^{^{\prime }})\,e^{i\vec{k}^{^{\prime }}\cdot \vec{r%
}},  \label{delta epsr}
\end{equation}
where 
\begin{equation}
\delta \vec{\varepsilon}(\vec{k}^{^{\prime }})=\frac{k_z^{^{\prime }}}%
K\delta \varepsilon _r\,\hat{e}_r\left( 2\pi \right) ^2\delta (\vec{k}%
_{\perp }^{^{\prime }}-\vec{k}_{0\perp })+\,\delta \vec{\varepsilon}_d(\vec{k%
}^{^{\prime }}).  \label{delta epsk}
\end{equation}
To calculate $\delta \vec{\varepsilon}\left( \vec{r}\right) $ we can proceed
exactly as in the previous section (3.3): $\delta \vec{\varepsilon}\left( 
\vec{r}\right) $ is the field radiated by a current $\delta \vec{J}(\vec{r})$
in the presence of the medium $\chi _s\left( \vec{r}\right) $. The current 
\begin{equation}
\delta \vec{J}(\vec{r})=\frac{-i\omega }{4\pi }\delta \chi (\vec{r})\vec{E}%
\left( \vec{r}\right) ,  \label{deltaJ}
\end{equation}
originates in the polarization of the roughness $\delta \chi (\vec{r})$ by
the total electric field $\vec{E}(\vec{r})$ due to the incident and all
scattered waves, including those generated by the roughness itself. Thus,
the challenge here is that the field $\vec{E}\left( \vec{r}\right) $ is
itself unknown; an approximation for it must be obtained as part of our
solution.

We can exploit the arbitrariness in the separation of $\chi \left( \vec{r}%
\right) $ into $\chi _s\left( \vec{r}\right) $ plus $\delta \chi \left( \vec{%
r}\right) $ to suggest a self-consistent approximation for $\vec{E}$.
Suppose we choose $z_0$ positive and considerably larger than the roughness $%
\sigma $ (see fig. \ref{fig3}). Then $\delta \chi \left( \vec{r}\right) $
represents a fictitious overlayer that extends well into the vacuum; the
sign of $\delta \chi \left( \vec{r}\right) $ is opposite to that of $\chi
_s\left( \vec{r}\right) $ and in the vicinity of $z_0$ they completely
cancel out. The field $\delta \vec{\varepsilon}(\vec{k})$ in a direction $%
\vec{k}$ with polarization $\hat{e}$ is given by eq.(\ref{Erad1}) 
\begin{equation}
\widehat{e}\cdot \delta \vec{\varepsilon}(\vec{k})=\frac{iK}2\int dv\,\delta
\chi (\vec{r})\vec{E}\left( \vec{r}\right) \cdot \vec{\varepsilon}_{ct}\,e^{i%
\vec{k}_{ct}\cdot \vec{r}}\,,  \label{dek}
\end{equation}
where the connecting field is precisely as in eqs.(\ref{ConnE})-(\ref{ect})
except for phase shifts due to the reflecting surface being at $z_0$, 
\begin{equation}
\vec{\varepsilon}_{cr}=e^{-2iqz_0}r_s\hat{e}_{cr}\quad \text{and}\quad \vec{%
\varepsilon}_{ct}=e^{i(\bar{q}-q)z_0}\,t_s\hat{e}_{ct}.  \label{epsct}
\end{equation}

\FRAME{ftbpFU}{3.9998in}{2.5002in}{0pt}{\Qcb{The problem of scattering by a
rough surface can be tackled using the ART by adding a fictitious overlayer $%
\delta \chi $. }}{\Qlb{fig3}}{afrp-f3.wmf}{\special{language "Scientific
Word";type "GRAPHIC";maintain-aspect-ratio TRUE;display "USEDEF";valid_file
"F";width 3.9998in;height 2.5002in;depth 0pt;original-width
306.8125pt;original-height 189.4375pt;cropleft "0";croptop "1";cropright
"1";cropbottom "0";filename
'C:/A-PAPERS/RecipThm/submit/figure3.eps';file-properties "XNPEU";}}

The reason behind the somewhat surprising choice for $z_0$ will now become
clear: slightly above $z_0$, in vacuum, the exact field is 
\begin{equation}
\vec{E}\left( \vec{r}\right) =\hat{e}_0e^{i\vec{k}_0\cdot \vec{r}}+\vec{%
\varepsilon}(\vec{r})=\hat{e}_0e^{i\vec{k}_0\cdot \vec{r}}+\vec{\varepsilon}%
_s(\vec{r})+\delta \vec{\varepsilon}\left( \vec{r}\right) ,  \label{exact E}
\end{equation}
but slightly below $z_0$ and, in fact, over all of the extension occupied by 
$\delta \chi \left( \vec{r}\right) $, we are also in vacuum ($\delta \chi
\left( \vec{r}\right) $ and $\chi _s\left( \vec{r}\right) $ cancel each
other) and therefore $\vec{E}\left( \vec{r}\right) $ is given by the same
expression (\ref{exact E}). The last term $\delta \vec{\varepsilon}\left( 
\vec{r}\right) $, given by (\ref{delta eps}), includes some weak diffusely
scattered and evanescent waves $\delta \vec{\varepsilon}_d(\vec{r})$. Our
approximation consists of neglecting them. Therefore, 
\begin{equation}
\vec{E}\left( \vec{r}\right) \approx \hat{e}_0\,e^{i\vec{k}_0\cdot \vec{r}}+r%
\hat{e}_re^{i\vec{k}_r\cdot \vec{r}},
\end{equation}
where the specular reflections by $\chi _s\left( \vec{r}\right) $ and $%
\delta \chi \left( \vec{r}\right) $ have been combined into the single, and
still unknown, reflection coefficient $r$, 
\begin{equation}
r=r_se^{-2iqz_0}+\delta \varepsilon _r.
\end{equation}
Substituting into eq.(\ref{dek}) yields

\begin{equation}
\widehat{e}\cdot \delta \vec{\varepsilon}(\vec{k})=-\frac{iK\chi _0}2\int
dx\,dy\int_{\zeta (x,y)}^{z_0}dz\,\left[ \hat{e}_0\,e^{i\vec{k}_0\cdot \vec{r%
}}+r\hat{e}_re^{i\vec{k}_r\cdot \vec{r}}\right] \cdot \vec{\varepsilon}%
_{ct}\,e^{i\vec{k}_{ct}\cdot \vec{r}}\,.  \label{edek}
\end{equation}
From now on we focus our attention on the specularly reflected component;
let $\hat{e}=\hat{e}_c=\hat{e}_r$, $\hat{e}_{cr}=\hat{e}_0$, $\vec{k}=\vec{k}%
_r=-\vec{k}_c$. Substituting eq.(\ref{delta epsk}) into the left hand side ($%
l.h.s.$), using $\left( 2\pi \right) ^2\delta (k_{\perp }-k_{0\perp
})=\left( 2\pi \right) ^2\delta (0)=\int dx\,dy$ we get 
\begin{equation}
l.h.s.=\frac qK\delta \varepsilon _r\left( 2\pi \right) ^2\delta (0)=\frac
qK\left( r-r_se^{-2iqz_0}\right) \int dx\,dy.  \label{lhs}
\end{equation}
This shows that the unknown reflection coefficient $r$ we want to calculate
appears in both the left and the right hand sides of (\ref{edek}), as part
of the radiated field and also as part of the field that induces the source;
eq.(\ref{edek}) permits a self-consistent calculation of $r$.

The integral over $dz$ in the right hand side of eq.(\ref{edek}) is
elementary and the remaining integral over $dx\,dy$ is performed using the
identity 
\begin{equation}
\frac{\int dx\,dy\,e^{-iQ\zeta (x,y)}}{\int dx\,dy}=\langle e^{-iQ\zeta
}\rangle =e^{-Q^2\sigma ^2/2},
\end{equation}
where $\zeta $ is a Gaussian random variable with zero mean, $\left\langle
\zeta \right\rangle =0$, and variance $\left\langle \zeta ^2\right\rangle
=\sigma ^2$. The right hand side ($r.h.s.$) of eq.(\ref{edek}) becomes 
\[
r.h.s.=\frac{K\chi _0}2\left( \int dx\,dy\right) \left\{ \frac{\hat{e}%
_0\cdot \vec{\varepsilon}_{ct}}{q+\bar{q}}\,\left[ \,e^{-i(q+\bar{q}%
)z_0}-e^{-(q+\bar{q})^2\sigma ^2/2}\right] \right. 
\]
\begin{equation}
-\left. \,r\,\frac{\hat{e}_r\cdot \vec{\varepsilon}_{ct}}{q-\bar{q}}\,\left[
\,e^{i(q-\bar{q})z_0}-e^{-(q-\bar{q})^2\sigma ^2/2}\right] \right\} ,
\label{rhs}
\end{equation}
which can be further rewritten by substituting $\vec{\varepsilon}_{ct}$ as
given by eq.(\ref{epsct}), and using $\bar{q}^2-q^2=K^2\chi _0$, and 
\begin{equation}
\frac{\hat{e}_0\cdot \hat{e}_{ct}}{\hat{e}\cdot \hat{e}_{ct}}=\frac{\hat{e}%
_{cr}\cdot \hat{e}_{ct}}{\hat{e}\cdot \hat{e}_{ct}}=\frac{\hat{e}_r\cdot 
\hat{e}_t}{\hat{e}_0\cdot \hat{e}_t}\quad \text{and}\quad \frac{\hat{e}%
_r\cdot \hat{e}_{ct}}{\hat{e}\cdot \hat{e}_{ct}}=1.  \label{ident pol}
\end{equation}
Finally, equating eq.(\ref{lhs}) to (\ref{rhs}) yields a self-consistent
approximation to $r$, 
\begin{equation}
r=\frac{q-\bar{q}}{q+\bar{q}}\,\frac{\hat{e}_r\cdot \hat{e}_t}{\hat{e}%
_0\cdot \hat{e}_t}\,e^{-2q\bar{q}\sigma ^2}=r_s\,e^{-2q\bar{q}\sigma ^2}.
\end{equation}
This coincides exactly with the Nevot-Croce result for the polarization $%
\hat{e}_0=\hat{e}_t=$ $\hat{e}_r$ for which the ratio $\hat{e}_0\cdot \hat{e}%
_t/$ $\hat{e}_r\cdot \hat{e}_t$ is unity, and provides the correct
generalization to all polarizations. According to this approximation the
specular reflection coefficient $r$ has no polarization dependence beyond
that already implicit in the reflection coefficient $r_s$ for the ideal flat
step surface; the ``static Debye-Waller'' factor $exp(-2q\bar{q}\sigma ^2)$
is polarization independent.

Notice that any possible dependence on the arbitrary choice of $z_0$ has
cancelled out.

\subsection{Reflection by smoothly graded surfaces}

The problem of scattering by a smoothly graded interface is similar and
somewhat simpler. Here the susceptibility $\chi (z)$ depends only on the
normal coordinate $z$ and not on the transverse coordinates $x$ and $y$.
This implies that the tangential component of momentum is conserved in the
scattering; there are no diffuse waves, there is only specular scattering.

As before, it is convenient to separate $\chi \left( z\right) $ into 
\begin{equation}
\chi \left( z\right) =\chi _s\left( z\right) +\delta \chi \left( z\right) ,
\end{equation}
where $\chi _s\left( z\right) $ represents an ideally flat surface at $z_0$, 
\begin{equation}
\chi _s\left( z\right) =\left\{ 
\begin{array}{ll}
0 & \text{for\quad }z>z_0 \\ 
\chi _0 & \text{for\quad }z<z_0
\end{array}
\right.
\end{equation}
and $\delta \chi \left( z\right) $ is an overlayer (see fig.\ref{fig4})
describing the smooth transition from bulk to vacuum.

\FRAME{ftbpFU}{4.4451in}{2.9101in}{0pt}{\Qcb{The problem of reflection by a
graded surface can be tackled using the ART by adding a fictitious overlayer 
$\delta \chi $. The hatched region shows the transition region from $\delta
\chi =0$ to $\delta \chi =-\chi _0.$ }}{\Qlb{fig4}}{afrt-f4.wmf}{%
\special{language "Scientific Word";type "GRAPHIC";maintain-aspect-ratio
TRUE;display "USEDEF";valid_file "F";width 4.4451in;height 2.9101in;depth
0pt;original-width 306.8125pt;original-height 199.125pt;cropleft "0";croptop
"1";cropright "1";cropbottom "0";filename
'C:/A-PAPERS/RecipThm/submit/figure4.eps';file-properties "XNPEU";}}

Let $\hat{e}_0\,e^{i\vec{k}_0\cdot \vec{r}}$ be the incident field. The
total scattered field $\vec{\varepsilon}(\vec{r})$, eq.(\ref{scattE}), 
\begin{equation}
\vec{\varepsilon}(\vec{r})=\vec{\varepsilon}_s(\vec{r})+\delta \vec{%
\varepsilon}\left( \vec{r}\right) .  \label{scattE gr}
\end{equation}
includes the wave $\vec{\varepsilon}_s(\vec{r})$ reflected by the step $\chi
_s\left( z\right) $, eq.(\ref{epss}), plus waves $\delta \vec{\varepsilon}%
\left( \vec{r}\right) $ scattered by the overlayer $\delta \chi \left(
z\right) $. While diffusely scattered waves are not present in $\delta \vec{%
\varepsilon}\left( \vec{r}\right) $, faint evanescent waves could be; these
are weak near field effects and we neglect them. Thus 
\begin{equation}
\delta \vec{\varepsilon}\left( \vec{r}\right) =\delta \varepsilon _r\,\hat{e}%
_re^{i\vec{k}_r\cdot \vec{r}},  \label{delta eps gr}
\end{equation}
and the Fourier expansion, eq.(\ref{delta epsr}), and transform $\delta \vec{%
\varepsilon}(\vec{k})$, eq.(\ref{delta epsk}), remain otherwise unchanged.
Once again, $\delta \vec{\varepsilon}\left( \vec{r}\right) $ is radiated by
a current 
\begin{equation}
\delta \vec{J}(\vec{r})=\frac{-i\omega }{4\pi }\delta \chi (z)\vec{E}\left( 
\vec{r}\right) ,  \label{deltaJ gr}
\end{equation}
where the field $\vec{E}(\vec{r})$ includes the incident and the unknown
reflected waves; $\vec{E}\left( \vec{r}\right) $ must be self-consistently
obtained as part of the solution. \smallskip Then the ART, in the form of
eq.(\ref{Erad1}), gives the field $\delta \vec{\varepsilon}(\vec{k})$ in a
direction $\vec{k}$ with polarization $\hat{e}$ as 
\begin{equation}
\widehat{e}\cdot \delta \vec{\varepsilon}(\vec{k})=\frac{iK}2\int dv\,\delta
\chi (z)\vec{E}\left( \vec{r}\right) \cdot \vec{\varepsilon}_{ct}\,e^{i\vec{k%
}_{ct}\cdot \vec{r}}\,,  \label{dek gr}
\end{equation}
with the same connecting field given back in eq.(\ref{epsct}).

The approximation we use for $\vec{E}\left( \vec{r}\right) $ is the same as
in last section. The arbitrariness of $z_0$ can be exploited by choosing it
large enough that the overlayer extends well into the vacuum. Near $z_0$ the
overlayer and the sharp step $\chi _s\left( \vec{r}\right) $ cancel each
other out; slightly above $z_0$, in vacuum, the field is 
\begin{equation}
\vec{E}\left( \vec{r}\right) =\hat{e}_0\,e^{i\vec{k}_0\cdot \vec{r}}+r\hat{e}%
_re^{i\vec{k}_r\cdot \vec{r}},  \label{exact E gr}
\end{equation}
where $r$ is the unknown reflection coefficient we want to calculate, 
\begin{equation}
r=r_se^{-2iqz_0}+\delta \varepsilon _r.
\end{equation}
Slightly below $z_0$ and over most of the extension occupied by $\delta \chi
\left( z\right) $ we are also in vacuum (provided the bulk to vacuum
transition is not too gradual) and we approximate $\vec{E}\left( \vec{r}%
\right) $ by the same expression, eq.(\ref{exact E gr}). Substituting into
eq.(\ref{dek gr}) yields an equation for $r$, 
\[
\frac qK\left( r-r_se^{-2iqz_0}\right) \left( 2\pi \right) ^2\delta
(k_{\perp }-k_{0\perp })= 
\]

\begin{equation}
=\frac{iK}2\int_{-\infty }^{z_0}dz\,\delta \chi (z)\int dx\,dy\,\left[ \hat{e%
}_0\,e^{i\vec{k}_0\cdot \vec{r}}+r\hat{e}_re^{i\vec{k}_r\cdot \vec{r}%
}\right] \cdot \vec{\varepsilon}_{ct}\,e^{i\vec{k}_{ct}\cdot \vec{r}}\,.
\label{ edek gr}
\end{equation}
The integral over $dx\,dy$ yields a delta function, $\left( 2\pi \right)
^2\delta (k_{\perp }-k_{0\perp })$, and we can substitute $\hat{e}=\hat{e}_c=%
\hat{e}_r$, $\hat{e}_{cr}=\hat{e}_0$, $\vec{k}=\vec{k}_r=-\vec{k}_c$. The
integral over $z$ is conveniently expressed as 
\begin{equation}
\int_{-\infty }^{z_0}dz\,\delta \chi (z)\,e^{-iQz}=\frac{\chi _0}{iQ}[%
\,e^{-iQz_0}+\frac{\chi ^{\prime }(Q)}{\chi _0}],  \label{integ Q}
\end{equation}
where $\chi ^{\prime }(Q)$ is the Fourier transform of $d\chi (z)/dz$, 
\begin{equation}
\chi ^{\prime }(Q)=\int_{-\infty }^{+\infty }dz\,\frac{d\chi (z)}{dz}%
\,e^{-iQz}  \label{chi prime}
\end{equation}
Eq.(\ref{integ Q}) is proved by integrating the left hand side by parts,
using $\delta \chi (z_0)\approx -\chi _0$, and $d\delta \chi (z)/dz=d\chi
(z)/dz$. Using $\bar{q}^2-q^2=K^2\chi _0$ and the identities in eq.(\ref
{ident pol}) the final result is 
\begin{equation}
r=r_s\,\frac{\chi ^{\prime }(\bar{q}+q)}{\chi ^{\prime }(\bar{q}-q)}.
\label{chi prime DW}
\end{equation}
Notice that any possible dependence on the arbitrary choice of $z_0$ has
cancelled out.

This coincides exactly with the scalar wave result \cite{Caticha95} and
provides the correct generalization to all polarizations. Within these
approximations the specular reflection coefficient $r$ has no polarization
dependence beyond that already implicit in the reflection coefficient $r_s$
for the ideal flat step surface; the ``static Debye-Waller'' factor is
polarization independent.

To conclude we mention some illustrative examples:

\noindent (a) The error-function profile 
\begin{equation}
\chi (z)=\frac{\chi _0}{\sqrt{2\pi \sigma ^2}}\int_{-\infty }^zdx\,\exp
-\left( \frac{x^2}{2\sigma ^2}\right) ,
\end{equation}
gives 
\begin{equation}
\frac{\chi ^{\prime }(\bar{q}+q)}{\chi ^{\prime }(\bar{q}-q)}=e^{-2q\bar{q}%
\sigma ^2},
\end{equation}
the same factor obtained in the previous section for a Gaussian rough
surface. This is as expected, the error function is the averaged profile for
the Gaussian rough surface.

\noindent (b) The Epstein (or Fermi distribution) profile \cite{Caticha95} 
\begin{equation}
\chi (z)=\frac{\chi _0}{1+e^{-z/\sigma }},
\end{equation}
gives 
\begin{equation}
\frac{\chi ^{\prime }(\bar{q}+q)}{\chi ^{\prime }(\bar{q}-q)}=\frac{\bar{q}+q%
}{\bar{q}-q}\,\frac{\sinh [\pi \sigma (q-\bar{q})]}{\sinh [\pi \sigma (q+%
\bar{q})]}.
\end{equation}

\noindent (c) The triangular profile 
\begin{equation}
\chi (z)=\left\{ 
\begin{array}{lll}
\chi _0 & \text{for} & z<-\sigma /2 \\ 
\chi _0\left( 1-2z/\sigma \right) & \text{for} & |z|<\sigma /2 \\ 
0 & \text{for} & z>\sigma /2
\end{array}
\right. ,
\end{equation}
gives 
\begin{equation}
\frac{\chi ^{\prime }(\bar{q}+q)}{\chi ^{\prime }(\bar{q}-q)}=\frac{q-\bar{q}%
}{q+\bar{q}}\,\frac{\sin [(q+\bar{q})\sigma /2]}{\sin [(q-\bar{q})\sigma /2]}%
.
\end{equation}

The reliability of these approximations was studied in \cite{Caticha95} in
the case of scalar waves. There is no reason to expect any difference from
the conclusions reached there: the ``static Debye-Waller'' in eq.(\ref{chi
prime DW}) provides a remarkably good approximation for the intensities
reflected by interfaces of arbitrary grading profile even for transition
regions that are quite wide ($\sigma $ as large as several nanometers). The
phase of the reflected waves is however more sensitive; eq.(\ref{chi prime
DW}) provides a good approximation for more abrupt transitions ($\sigma $ of
the order of 1 \emph{nm} or less).

\section{Conclusion}

The main result of this work, eq.(\ref{art}), is an asymptotic form of the
reciprocity theorem which can be used as the basis for a practical method
for calculations. The theorem states that the field radiated in the presence
of a nontrivial medium, in a certain direction and with a given
polarization, is a suitable `component' of the radiating source. This
`component' is to be extracted by introducing an auxiliary `connecting'
field which contains the necessary information about the medium. The
practical advantage of the method lies in the simplifications achieved by
systematically avoiding unnecessary calculations; it thereby allows one to
tackle problems of increasing complexity.

In forthcoming papers we will further explore the application of the ART to
the study of the dynamical diffraction of radiation generated by sources
within a crystal, the so-called Kossel lines. Even this well explored topic
has not been exhausted. Of particular interest are situations where the
Bragg angle lies close to $\pi /2$ and the Kossel cones degenerate into
single beams \cite{Zambianchi99}, and situations where the source location
is revealed by the oscillatory `Pendell\"{o}sung' structure of the
diffraction pattern \cite{Sutter99}. Other applications will include a new
approach to thermal diffuse scattering under conditions of dynamical
diffraction \cite{Zambianchi99b}.

\textbf{Acknowledgments}. I am indebted to P. Zambianchi and E. Sutter for
valuable discussions.


\begin{thebibliography}{99}
\bibitem{footnote1}  This is a nontrivial statement; its justification is
given in the paragraph following eq. \ref{I2}.
\bibitem{footnote2}  To avoid factors of $4\pi $ elsewhere it is usual in
x-ray optics to define the susceptibility by $4\pi P=\chi E$.
\end{thebibliography}
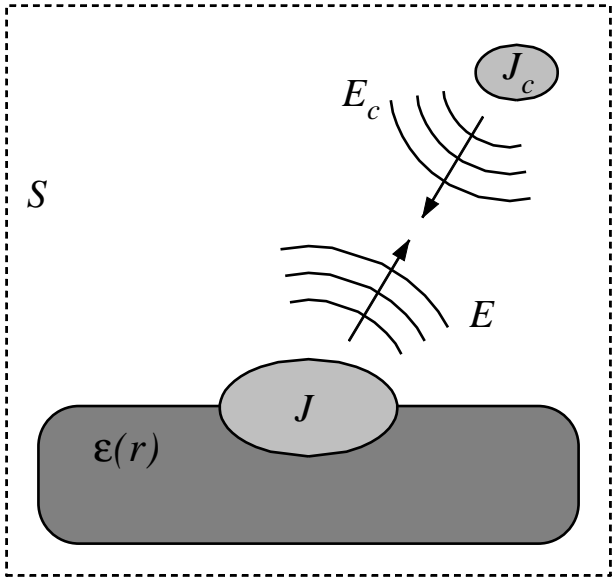
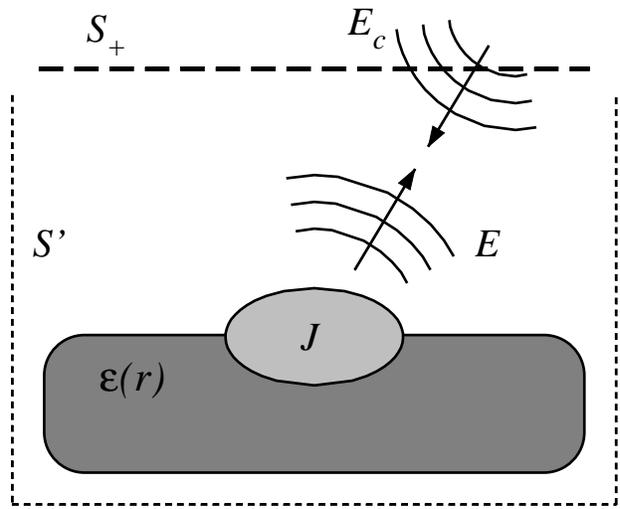
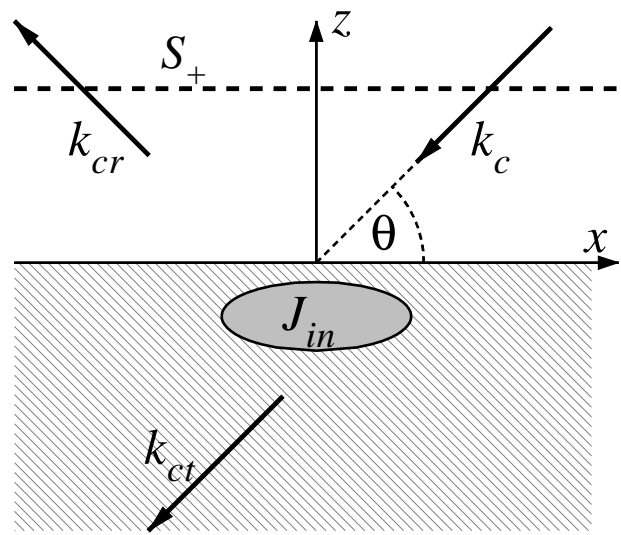
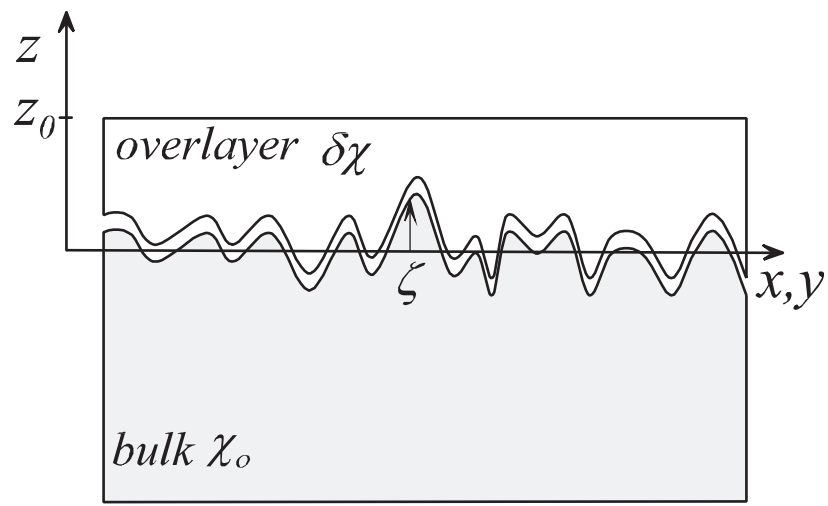
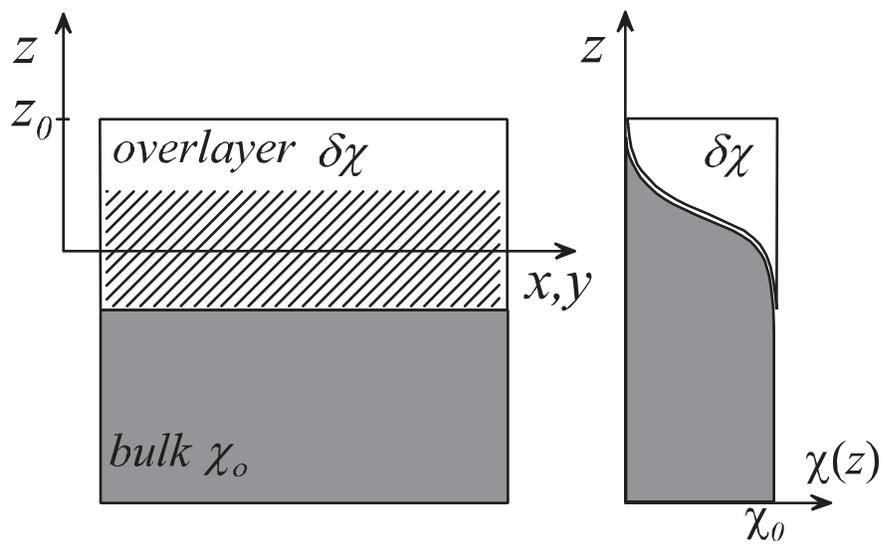
\end{document}